\documentclass{article}
\usepackage{spconf,amsmath,graphicx}
\usepackage{adjustbox}
\usepackage{hyperref}
\usepackage{makecell}
\usepackage{multirow}
\usepackage{amsmath}
\usepackage{float}
\usepackage{subfigure}
\usepackage[inkscapelatex=false]{svg}
\usepackage{booktabs}
\usepackage{cite}
\usepackage[hyphenbreaks]{breakurl}
\usepackage{xurl}

\hypersetup{
    colorlinks,
    breaklinks,
    urlcolor=black,
    linkcolor=black
}
\title{Towards general-purpose text-instruction-guided voice conversion}
\name{\begin{tabular}[c]{@{}c@{}c@{}c@{}c@{}c@{}}
      Chun-Yi Kuan$^{1}$, Chen-An Li$^{1}$, Tsu-Yuan Hsu$^{1}$, Tse-Yang Lin$^{1}$, Ho-Lam Chung$^{1}$, Kai-Wei Chang$^{1}$ \\ 
      Shuo-yiin Chang$^{2}$, Hung-yi Lee$^{1}$
      \end{tabular}
      }

\address{\begin{tabular}[c]{@{}c@{}}
     $^{1}$National Taiwan University, Taiwan,
     $^{2}$Google, USA
\end{tabular}}

\copyrightnotice{979-8-3503-0689-7/23/\$31.00~\copyright2023 IEEE}
\begin{document}
\ninept
\maketitle
\begin{abstract}
This paper introduces a novel voice conversion (VC) model, guided by text instructions such as ``articulate slowly with a deep tone'' or ``speak in a cheerful boyish voice''. 
Unlike traditional methods that rely on reference utterances to determine the attributes of the converted speech, using text instruction adds versatility and specificity to voice conversion. 
The proposed VC model is a neural codec language model which processes a sequence of discrete codes, resulting in the code sequence of converted speech.  
It utilizes text instructions as style prompts to modify the prosody and emotional information of the given speech. 
In contrast to previous approaches, which often rely on employing separate encoders like prosody and content encoders to handle different aspects of the source speech, our model handles various information of speech in an end-to-end manner. 
Experiments have demonstrated the impressive capabilities of our model in comprehending instructions and delivering reasonable results\footnote[1]{\label{fn:asru2023-demo}{\scriptsize\url{https://text-guided-vc.github.io/asru2023_demo}}}.
% \footnote[0]{See \href{https://text-guided-vc.github.io/asru2023_demo/}{\color{blue} here} for demonstrations of our work.}.
% \footnote[0]{See \href{https://asru2023taipei.github.io/demo}{https://asru2023taipei.github.io/demo} for demonstrations of our work.}


% See here\footnote[0]{demo link} for demos of our work.

%Utilizing text descriptions as prompts to guide the generation of texts, images, and speech synthesis has garnered interest recently. 
% This study explores the use of text instructions to guide voice conversion, a departure from conventional methods that rely on reference utterances to determine the attributes of the converted speech. We present a novel voice conversion model guided by text instructions, such as "articulate it slowly with a deep tone" or "speak in a cheerful boyish voice," thereby expanding the versatility and specificity of voice conversion.
% We introduce a language model approach for this task. 
%Particularly, we train a neural codec language model by offline exploiting discrete codes derived from the neural codec model, and consider voice conversion to be a conditional language modeling task. During training, we utilize natural language instructions as style prompts to modify the prosody and emotional information of the source speech. 
%In contrast to previous approaches, which often rely on employing separate encoders like prosody and content encoders to handle different aspects of the source speech, our model can handle various information of speech in an end-to-end manner. 
%Experiments have demonstrated the impressive capabilities of our model in comprehending instructions and delivering satisfactory results. See here\footnote[0]{demo link} for demos of our work.

\end{abstract}

%
%\begin{keywords}
%
%\end{keywords}
%

\section{Introduction}
\label{sec:intro}
In the field of natural language processing (NLP), it has become crucial to enable language models to comprehend human intentions through instructions. As a result, several instruction-tuned language models have been proposed, such as InstructGPT \cite{ouyang2022training}, aiming to enhance language model's understanding by fine-tuning the model with natural language instructions.

In the domain of speech processing, efforts have also been made to incorporate textual information into speech or audio generation \cite{rubenstein2023audiopalm, 
le2023voicebox, ghosal2023text, 
huang2023make, kharitonov2023speak, 
borsos2023soundstorm, wang2023neural, 
liu2023audioldm, zhang2023speak, 
copet2023simple, wang2023viola, liu2023promptstyle}. Approaches like PromptTTS \cite{guo2023prompttts} and AudioGen \cite{kreuk2023audiogen} can synthesize novel speech based on textual descriptions, enabling tasks such as prompted text-to-speech or text-to-audio. However, to the best of our knowledge, there is a lack of research exploring the potential of instruction-guided speech-to-speech tasks.

\begin{figure*}[t]
    \centering
    \includegraphics[width=0.75\textwidth]{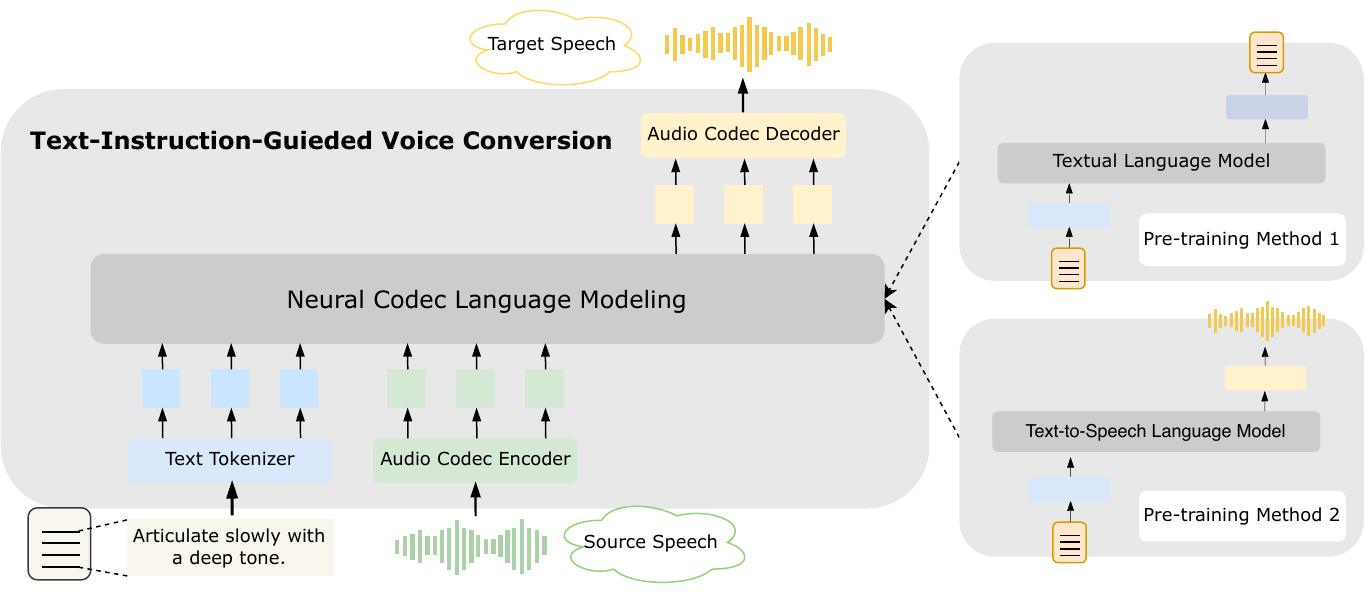}
\textbf{}
    % \caption{The overview of our study. Taking a text-instruction and source speech as prompts, our work is able to generate the discrete audio codes of target speech, which can be decompressed to the waveform of target speech.}
    
    % \caption{The overview of our framework. Conditioned on the given textual instruction and source speech, our model is able to generate sequences of discrete audio codes, which can be decompressed to the waveform of target speech. 

    \caption{The overview of our framework. Conditioned on the given textual instruction and source speech, our model is able to generate sequences of discrete audio codes. Subsequently, we decode these sequences of codes to the corresponding target speech by audio codec decoder.
    In addition, we propose two pre-training methods. Method 1 involves fine-tuning the textual language model. Method 2 involves fine-tuning the text-to-speech pre-trained language model.}

    % Method 2 involves pre-training the textual language model on the text-to-speech task before fine-tuning it
    
    \label{Overview}
    
\end{figure*}

In this paper, we propose a novel framework for text-instruction-guided voice conversion, which is a groundbreaking task.  The detailed architecture of our framework is illustrated in Fig \ref{Overview}. Guided by text instructions, our model is capable of transforming the given source speech into the style that aligns with the provided instructions. For instance, conditioned on instructions like ``speak with a lively and enthusiastic boyish tone'' or ``give me a mature tone in a deep and somber voice'',  the transformed speech can effectively emulate the desired speaking styles.
%the converted speech sounds like ``a lively and enthusiastic boyish voice'' and ``a deep and somber voice that conveys maturity'' respectively. 

% For instance, conditioned on instructions like ``speak with a lively and enthusiastic boyish tone'' and ``give me a mature tone in a deep and somber voice'', the converted speech sounds like ``an energetic and youthful voice reflecting enthusiasm and liveliness'' and ``a deep voice that conveys maturity and sorrow'' respectively. 
% The converted speech, upon instruction, resonates with an ebullient and enthusiastic boyish tone in one instance and exudes a mature, deep, and somber vocal characteristic in the other.
%Our framework can handle versatile textual instructions without any constraints on the specific format or type of the instructions. 
The primary objective of this research is to create a  voice conversion model capable of processing a wide range of textual instructions. 
Voice conversion traditionally relies on reference-based style control \cite{sisman2020overview, lorenzo2018voice, mohammadi2017overview, zhao2020voice, zhou2022emotional}, 
where a style reference speech clip is utilized to guide the style of the conversion. However, this method can present several challenges. For instance, obtaining an appropriate style reference can be difficult, and it may not encompass all the nuanced stylistic elements desired in the output speech.
Text-based style control offers an innovative solution to these challenges, with several distinct advantages outlined below.

% \begin{enumerate}

% \item {Flexibility. Text-based style control offers enhanced flexibility and facilitates a more diverse representation of style by accommodating any form of human language as a style descriptor. This ``form'' is not confined to simple or predefined commands. It encompasses everything from single-word descriptors to complex sentences, idiomatic expressions, and highly detailed descriptions. For instance, beyond basic instructions like ``speak slowly'' or ``use a high-pitched voice'', the model could also understand and implement nuanced instructions such as ``speak as if you're telling a bedtime story to a child'' or ``adopt the tone of a news anchor delivering breaking news''. This unconstrained approach obviates the need for a specific format or type and covers a broader range of styles that might not be easily captured by a single reference speech.}

% \item{Ease of use. Text-based style instructions can be more intuitive and natural to use, especially for non-expert users. Users can simply write down the desired style in text, which can be more accessible and straightforward than finding or creating an appropriate speech reference.}

% \item{Potential for richness and nuance. Language is a powerful tool that can express a wide range of concepts, emotions, and nuances. By leveraging the richness of language, text-based style control can potentially guide voice conversion in a more nuanced and detailed manner.}

% \end{enumerate}

\noindent \textbf{(1) Flexibility}. Text-based style control offers enhanced flexibility and facilitates a more diverse representation of style by accommodating any form of human language as a style descriptor. This ``form'' is not confined to simple or predefined commands. It encompasses everything from single-word descriptors to complex sentences, idiomatic expressions, and highly detailed descriptions. For instance, beyond basic instructions like ``speak slowly'' or ``use a high-pitched voice'', the model could also understand and implement nuanced instructions such as ``speak as if you're telling a bedtime story to a child'' or ``adopt the tone of a news anchor delivering breaking news''. This unconstrained approach obviates the need for a specific format or type and covers a broader range of styles that might not be easily captured by a single reference speech. 

\noindent \textbf{(2) Ease of use}. Text-based style instructions can be more intuitive and natural to use, especially for non-expert users. Users can simply write down the desired style in text, which can be more accessible and straightforward than finding or creating an appropriate speech reference.

\noindent \textbf{(3) Potential for richness and nuance}. Language is a powerful tool that can express a wide range of concepts, emotions, and nuances. By leveraging the richness of language, text-based style control can potentially guide voice conversion in a more nuanced and detailed manner.
%%

% \item {Adaptability. Unlike reference-based control, text instructions can easily adapt to different speech contents and contexts, providing more consistent style conversion results.}

% \end{enumerate}
% In summary, text-based style control provides a more flexible, intuitive, and potentially richer and more nuanced approach for voice conversion compared to traditional reference-based style control. 
% However, further research is needed to explore and optimize its applications.

Our proposed VC model is a neural codec language model which processes the discrete code sequence of the source speech encoded by the encoder of EnCodec \cite{defossez2022highfi} and results in the code sequence of converted speech. 
% The detailed architecture is depicted in Fig \ref{Overview}.
The detailed schematic of our method is presented in Fig \ref{Overview}.
The novelty and uniqueness of our proposed approach lie in two aspects. One is that we view voice conversion as a neural codec language modeling task, while the other is that we experiment with various model configurations, including utilizing pre-trained textual model, textual model pre-trained on text-to-speech, and even training from scratch.
As SpeechLMs \cite{hassid2023textually} garner benefits from initializing with textual LMs, we likewise observe that pre-training on a textual model markedly enhances its capacity to interpret textual instructions.

% We find that pre-training on textual model benefits greatly from the ability to comprehend textual instructions.
Moreover, due to the current lack of suitable datasets for training text-instruction-guided voice conversion models, which have to include source speech samples, target speech samples, and corresponding textual instructions, we take the initiative to establish two distinct datasets by means of the audio processing tool SoX \cite{sox} and PromptSpeech dataset \cite{guo2023prompttts} for our work. We refer to them as the Signal Processing Effect Dataset and InstructSpeech Dataset respectively.

%To ensure an objective assessment of our results, we employ human evaluation with two distinct criteria. 
%The first requests participants to gauge whether the effects displayed in the target speech aligned with the textual descriptions in the instructions, with the source speech as a reference, while disregarding speech quality. 
%The second criterion involves assessing solely the quality of the target speech, overlooking its effects. 
The subjective and objective evaluations reveal that our model possesses a remarkably compelling text comprehension capacity, successfully transforming and integrating the corresponding styles from the instructions into the target speech. 
%However, comparatively, the quality of the speech did not shine as brightly. As such, our future efforts will be geared towards enhancing speech quality.
% Our proposed method stands out for its uniqueness in two aspects.
Finally, to access the model's generalization and robustness capabilities, we conduct experiments to evaluate our model on out-of-domain instructions that are never seen during training. We accomplish unseen instructions by amalgamating distinct types of instructions, one exclusively appearing in the Signal Processing Effect Dataset and the other solely in the InstructSpeech Dataset.
The experimental results underscore the robust capability of the model to generalize effectively. Additionally, we delve into the impact of adverbs of varying degrees within the instructions. The findings suggest that our model is adept at discerning the subtle differences among adverbs and effectively utilizes this knowledge to influence the final converted speech appropriately.

\section{Related Work}
\label{sec:related_work}

\subsection{Neural Codec Model}
\label{ssec:neural_codec_model}

Neural audio codecs \cite{defossez2022highfi, zeghidour2021soundstream, wu2023audiodec} have offered significant benefits for the compression and transmission of audio data across networks. Furthermore, AudioLM \cite{borsos2023audiolm} generates audio as a language modeling task by utilizing hierarchical modeling of semantic and acoustic tokens, which achieved success in works such as SPEAR-TTS \cite{kharitonov2023speak} and SoundStorm \cite{borsos2023soundstorm}. In our work, we follow VALL-E \cite{wang2023neural} to utilize neural codec models to convert speech into discrete tokens. VALL-E introduces a prompt-based neural codec language model approach for TTS with audio codecs of EnCodec as intermediate representations. By encoding the speech waveform into discrete acoustic codes, these models enable the reconstruction of high-quality waveforms, even for speakers who were not present in the training data. 

In contrast to VALL-E, which is a TTS system that focuses on synthesizing personalized speech using a 3-second recording of an unseen speaker as an acoustic prompt, our work aims to propose a novel voice conversion model that can be guided by text instructions to achieve desired styles.

% IF NO SPACE LEFT, REMOVE THIS PARAGRAPH.
% Selecting neural codec models as the compression method  brings several advantages over alternative quantization approaches. Firstly, neural codec models excel in preserving abundant speaker information and acoustic characteristics, ensuring that the reconstructed speech maintains the speaker's identity more effectively compared to discrete codes derived from HuBERT \cite{hsu2021hubert}. This is crucial in our tasks where maintaining the speaker's distinctive traits is desirable. Secondly, the presence of an off-the-shelf codec decoder simplifies the process of converting discrete tokens back into a waveform. This eliminates the need for additional training efforts, such as vocoder training.

%By leveraging the capabilities of neural codec models, we achieve effective compression and reconstruction of speech while simultaneously preserving important speaker-related information and streamlining the decoding process. These inherent advantages make audio codec models a highly appealing and suitable choice for our specific task.

\subsection{Textual-guided speech generation}
\label{ssec:textual_guided_speech_generation}

% Recently, prompt-based methods in natural language processing have been increasingly used to control generated speech characteristics such as emotion and prosody. 
Recently, prompt-based methods have been explored in speech processing \cite{chang2022speechprompt, wu2023speechgen}. A notable research direction is to adopt text prompts for controlling speech characteristics, such as emotion and prosody, in speech generation \cite{guo2023prompttts, yang2023instructtts, liu2023promptstyle}. 
{AudioGen \cite{kreuk2023audiogen}}
% , an autoregressive model, 
generates audio samples from text inputs by encoding raw audio into discrete tokens with a neural audio compression model, then processing these tokens with a transformer-decoder language model considering the text inputs. 
% A text encoder model captures the semantic and contextual information of text, which is crucial for the resultant audio's desired attributes.
{PromptTTS \cite{guo2023prompttts}} uses text prompts for high-quality speech synthesis, aligning with style and content descriptors. Its architecture includes a style encoder extracting emotion and prosody, a content encoder for content-related details, and a speech decoder combining these representations to create speech waveforms. Similarly, {InstructTTS \cite{yang2023instructtts}} also employs prompts but without constraints on style descriptions, allowing natural language inputs for training.

In contrast to the aforementioned studies that focus on text-to-audio or text-to-speech generation,
our research proposes a text-guided speech-to-speech conversion model. Our approach utilizes textual instruction in any form of human language, unconfined by specific formats.
This offers intuitive use, harnesses language richness, and imbues voice conversion tasks with enhanced flexibility.

% Diverging from prior research focusing on text-to-audio or text-to-speech generation, our study introduces a text-guided speech-to-speech conversion model. We utilize text instructions in any human language form,  unconfined by specific formats, facilitating intuitive use and capitalizing on the richness of language to imbues voice conversion tasks with enhanced flexibility.

% In contrast to the aforementioned studies that focus on generating audio or speech from text, our study proposes a speech-to-speech conversion model where text is employed to guide the generation of speech. Text instruction can adopt any form of human language, unrestricted by specific formats or types. This not only affords intuitive and natural usage but also benefits from the richness of language, thereby infusing the voice conversion task with increased flexibility.

\subsection{Style conversion}
\label{ssec:style_conversion}

% Style conversion aims to transform a given speech into a customized speech based on a specific style. Many supervised and unsupervised algorithms have been developed to facilitate the conversion of speech styles.
Style conversion seeks to modify a given speech according to a specific style, with numerous supervised and unsupervised algorithms developed for this transformation.
To overcome the problem of emotional conversion, \cite{aihara2012gmm} proposed the use of a Gaussian Mixture Model (GMM) for jointly modeling the prosodic attributes of both the source and target, and \cite{inanoglu2009data} also proposed a comprehensive framework that integrates Hidden Markov Model (HMM), Gaussian Mixture Model (GMM), and a fundamental frequency segment selection method. Due to the rapid progress of deep net, research related to style conversion has also experienced a significant leap in advancement. In the case of \cite{gao19b_interspeech}, a speech-to-speech emotional conversion model is proposed using style transfer auto-encoders. Recent research also makes many efforts to address this issue. For example, \cite{kreuk-etal-2022-textless} leverage discrete and decomposed representations for speech
emotion conversion, and 
\cite{zhou20_odyssey, rizos2020stargan} employed GAN-based methods to achieve prosody conversion and emotional speech conversion respectively. 

% Voice conversion is also a topic that receives significant attention in many research studies, which aims to convert the source speaker voice into the target speaker voice while preserving the linguistic information. 
Voice conversion, a research focus aiming to transmute the source speaker's voice to the target's while maintaining the linguistic content, garners substantial attention in numerous studies.
Traditional VC methods concentrate on one-to-one voice conversion, where they align parallel data from the source and target speakers and acquire a frame-wise mapping \cite{sun2015voice, toda2007voice}. In recent studies, non-parallel data is predominantly utilized for more practical many-to-many voice conversion \cite{chou2019one, qian2019autovc, choi2021neural, casanova2022yourtts, nguyen2022nvc}. 
% This involves the widespread adoption of generative adversarial networks (GAN) \cite{kameoka2018stargan, kaneko2019stargan, kaneko2021maskcyclegan, pasini2019melgan, kaneko2017parallel, kaneko2018cyclegan, kaneko2019cyclegan, kaneko2020cyclegan, li2021starganv2, chou2018multi, fang2018high}, variational auto-encoders (VAE) \cite{hsu2016voice, hsu2017voice, huang2018voice, kameoka2019acvae, luo2020singing} and other auto-encoders \cite{qian2019autovc, qian2020f0}. 
This involves the widespread adoption of generative adversarial networks (GAN) such as StarGAN \cite{kameoka2018stargan, kaneko2019stargan, li2021starganv2}, CycleGAN \cite{kaneko2017parallel, fang2018high, kaneko2018cyclegan, kaneko2019cyclegan, kaneko2021maskcyclegan, kaneko2020cyclegan} and other variants \cite{chou2018multi, pasini2019melgan}. In addition, variational autoencoders (VAE) \cite{hsu2016voice, hsu2017voice, huang2018voice, kameoka2019acvae, luo2020singing} and other autoencoders \cite{qian2019autovc, qian2020f0} have also been employed.
\cite{lin2021s2vc, lin2021fragmentvc, van2022comparison, wang2022drvc, wang2022zero} have introduced the use of self-supervised learning (SSL) techniques or frameworks to facilitate the task of voice conversion. For instance, \cite{huang2022s3prl, huang2021any, choi2021neural, polyak2021speech} 
% utilize SSL representation as the linguistic content representation to solve the voice conversion task, which can excel in the disentanglement aspect.
leverage SSL representation for linguistic content in voice conversion, excelling in the disentanglement aspect.

However, speech conversions typically focus on individual aspects, and studies converting multiple styles like emotion, timbre, and prosody are rare. Our study aims to address this gap by employing text-based instructions to guide the model in multi-style conversions.
% AR-NAR Graph.
\begin{figure}[t]
    \centering
    \includegraphics[width=0.50\textwidth]{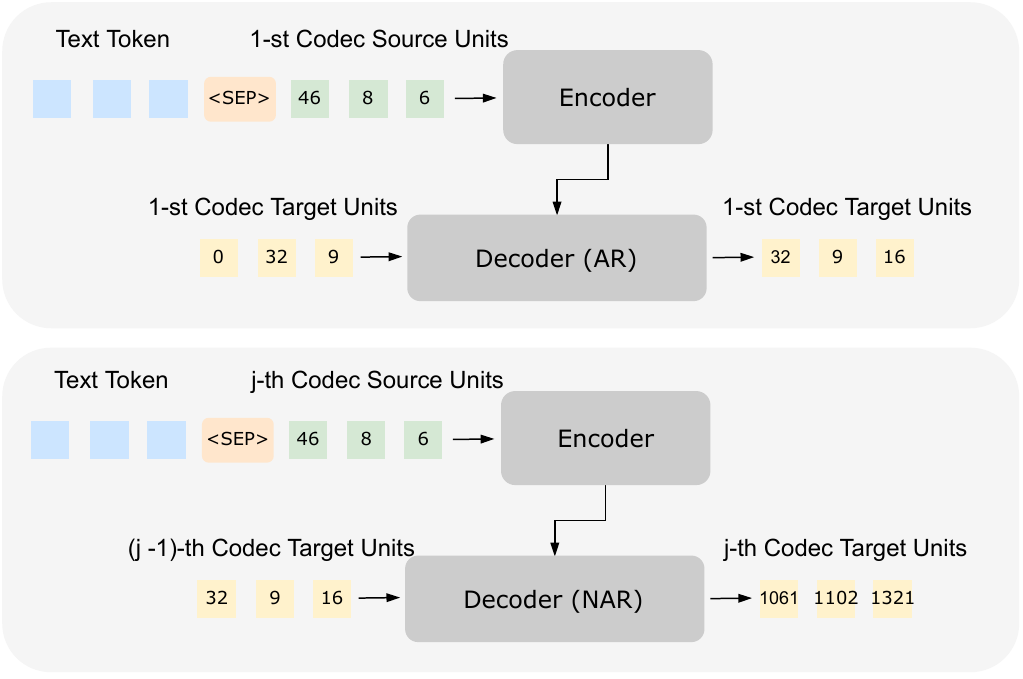}

    \caption{The architecture of autoregressive and non-autoregressive codec language model.}

    \label{AR-NAR-Architecture}
\end{figure}

\section{Method}
\label{sec:method}
In this section, we formulate our task: text instruction-guided voice conversion through conditional codec language modeling and introduce our proposed framework in detail.

\subsection{Problem formulation}
\label{ssec:problem_formulation}

% Instructed voice conversion
Text instruction-guided voice conversion aims to convert a speech sample $\mathbf{x} = (x_1, x_2, \dots, x_n)$ into a target speech sample $\mathbf{y} = (y_1, y_2, \dots, y_m)$, based on a user instruction $\mathbf{z} = (z_1, z_2, \dots, z_l)$.  $x_i$ and $y_i$ are the frames of the source speech and target speech respectively, and $z_i$ is the token of the instruction. $m$ and $n$ denote the number of frames, and $l$ denotes the number of tokens. In our approach, we consider the conversion process as conditional codec language modeling. Our framework consists of two key components: a neural audio codec and a conditional codec language model. The neural audio codec is responsible for encoding a speech sample into discrete acoustic codes and decoding the discrete acoustic codes back into speech. The conditional codec language model, on the other hand, leverages the encoded discrete acoustic codes along with the user instruction to generate the target acoustic codes. 

We formulate the conversion process into the following steps. First, we utilize the neural audio codec's acoustic quantizer, which consists of an encoder and several quantization layers, to encode each speech sample into discrete acoustic codes. We adopt EnCodec \cite{defossez2022highfi} as our neural audio codec. The quantized process can be formualted as $\mathbf{c}^{\mathbf{x}} = E_{codec}(\mathbf{x}), \mathbf{c}^{\mathbf{y}} = E_{codec}(\mathbf{y})$, where $\mathbf{c}^{\mathbf{x}} \in \mathcal{N}^{n' \times 8}$, $\mathbf{c}^{\mathbf{y}} \in \mathcal{N}^{m' \times 8}$. $n'$ and $m'$ denote the downsampled source utterance length and target utterance length respectively. Next, the conditional codec language model is conditioned on the encoded discrete acoustic codes and user instructions to produce the desired target acoustic codes, which can be seen as estimating $p(\mathbf{c}^{\mathbf{y}} | \mathbf{c}^{\mathbf{x}}, \mathbf{z}; \theta)$. $\theta$ denotes the parameter of the conditional codec language model. Finally, we can synthesize the waveform $\hat{\mathbf{y}}$ using the audio codec decoder, $\hat{\mathbf{y}} = D_{codec}(\mathbf{c}^{\hat{\mathbf{y}}})$, where $\mathbf{c}^{\hat{\mathbf{y}}}$ is the predicted target acoustic codes.

\subsection{Training: Conditional Codec Language Modeling}
\label{ssec:training}

In this work, we reach the functionality of the conditional codec language model by employing a hierarchical approach with two encoder-decoder codec language models. These two models utilize different modeling methods: autoregressive and non-autoregressive. We first apply autoregressive codec language modeling to generate the discrete codes from the first quantizer of the target speech sample. Following this, we exploit non-autoregressive codec language modeling to predict the remaining discrete codes. We can formulate our model as $\theta = \{ \theta_{ar}, \theta_{nar} \}$, where $\theta_{ar}$ and $\theta_{nar}$ denotes the parameter of the autoregressive codec language model and non-autoregressive codec language model respectively. We discuss the design of each modeling method in the following subsection.
\\

\noindent {\bf Autoregressive Codec Language Modeling}

In our approach, we employ autoregressive codec language modeling specifically for generating the discrete codes from the first quantizer of the target speech sample. 
This process can be regard as estimating $p(\mathbf{c}^{\mathbf{y}}_{:,1}|\mathbf{c}^{\mathbf{x}}_{:,1}, \mathbf{z}; \theta_{ar}) = \prod_{t=1}^{m'}p(\mathbf{c}^{\mathbf{y}}_{t,1}|\mathbf{c}^{\mathbf{y}}_{1:t-1,1}, \mathbf{c}^{\mathbf{x}}_{:,1}, \mathbf{z}; \theta_{ar})$, where $\mathbf{c}_{:,1}$ denotes the discrete codes from the first quantizer. We select autoregressive modeling due to the potential inconsistency in length between the source speech and the target speech. By utilizing autoregressive modeling, we can effectively handle variations in the length of the target speech during the conversion process. We implement the autoregressive codec language model by using transformer-based encoder-decoder architecture. The encoder input is the concatenation of $\mathbf{z}$ and $\mathbf{c}^{\mathbf{x}}_{:,1}$, and the decoder input is the right shift $\mathbf{c}^{\mathbf{y}}_{:,1}$. Illustrated in the upper part of Fig. \ref{AR-NAR-Architecture}.
\\

\noindent {\bf Non-Autoregressive Codec Language Modeling}

We employ non-autoregressive codec language modeling to generate the remaining discrete codes of the target speech sample. This can be view as estimating $p(\mathbf{c}^{\mathbf{y}}_{:,j}|\mathbf{c}^{\mathbf{y}}_{:,j-1}, \mathbf{c}^{\mathbf{x}}_{:,j}, \mathbf{z}; \theta_{nar})$, $j \in [2,8]$, where $\mathbf{c}_{:,j}$ denotes the discrete codes from the $j$-th quantizer, $\theta_{nar}$ denotes the parameter of the non-autoregressive codec language model. Using non-autoregressive modeling have two benefits. 
The pre-trained neural codec decoder $D_{codec}(\cdot)$ should take the same length of discrete codes as input, which means $|\mathbf{c}^{\mathbf{y}}_{:,j}| = |\mathbf{c}^{\mathbf{y}}_{:, i}|$, $\forall$ $i \neq j$. Employing a non-autoregressive model to predict the discrete codes can achieve this goal with little effort. Also, non-autoregressive modeling enhances the generating process efficiency. We take a transformer-based encoder-decoder model as our non-autoregressive codec language model, which is similar to the architecture of our autoregressive codec language model, except for the self-attention masks of the decoder. We adopt the concatenation of $\mathbf{c}^{\mathbf{x}}_{:,j}$ and $\mathbf{z}$ as the encoder input, and take $\mathbf{c}^{\mathbf{y}}_{:,j-1}$ as the decoder input. Illustrated in the lower part of Fig. \ref{AR-NAR-Architecture}.

% \subsection{Model Configuration}
%\subsection{Training setup}
\subsection{Pre-training}
\label{ssec:pre_training}

We investigate two ways to pre-train the AR model and the NAR model. 
The two pre-training methods are illustrated in Fig \ref{Overview}.
In the following discussion, we use the term ``Scratch'' to represent models that have not undergone any form of pre-training.
%Both the AR model and the NAR model have the following three training settings.
 %   \item \textbf{Train from scratch}. We utilize the same architecture of the above textual language model but without pre-trained weights. We refer to this setting as ``Scratch''.
% \begin{enumerate}

\noindent \textbf{(1) Fine-tune the textual language model}. The textual language model, like BART~\cite{lewis2020bart}, has the encoder-decoder architecture. We fine-tune our neural codec language model on the textual language model as our initialization for the voice conversion task. We refer to this setting as ``Text''.
    
\noindent \textbf{(2) Fine-tune the text-to-speech pre-trained language model}. 
In ``Text'',  the pre-trained model is pre-trained without exposure to speech data. 
To enhance the pre-training model, we proceed to further pre-train it on a text-to-speech task. 
In the text-to-speech task, the input is text, and the output is modified to a sequence of discrete codes. 
Subsequently, we decode this sequence of codes back to the corresponding speech. 
 %   We adopt the textual language model and utilize text-to-speech as the pre-trained task beforehand and followed by fine-tuning our neural codec language model from it. 
We refer to this setting as ``TTS''.

%    \item \textbf{Train from scratch}. We utilize the same architecture of the above textual language model but without pre-trained weights. We refer to this setting as ``Scratch''.
    
% \end{enumerate}
Both AR and the NAR models can have the three different aforementioned pre-training configurations: ``Scratch'', ``Text'', ``TTS'', resulting in a total of nine possible combinations of models.

\subsection{Inference}
\label{ssec:inference}

Given a source speech sample $\mathbf{x} = (x_1, x_2, \dots, x_n)$ and a text instruction $\mathbf{z} = (z_1, z_2, \dots, z_l)$, $n$ and $l$ denote the length of the utterance and text respectively. First, we utilize the neural audio codec's acoustic quantizer to encode each speech sample into discrete acoustic codes, $\mathbf{c}^{\mathbf{x}} = E_{codec}(\mathbf{x})$, where $\mathbf{c}^{\mathbf{x}} \in \mathcal{N}^{n' \times 8}$, and $n'$ denote the downsampled utterance length. Second, inference process can be regarded as sampling $\mathbf{c}^{\hat{\mathbf{y}}}$  $\sim$ $p(\mathbf{c}^{\mathbf{y}} | \mathbf{c}^{\mathbf{x}}, \mathbf{z} ; \theta)$ using the multinomial sampling strategy, where $\theta$ denotes the parameter of the conditional codec language model and $\mathbf{c}^{\hat{\mathbf{y}}}$ is the generated target sequence of discrete codes. Finally, we can synthesize the waveform $\hat{\mathbf{y}}$ using the pre-trained neural codec decoder, $\hat{\mathbf{y}} = D_{codec}(\mathbf{c}^{\hat{\mathbf{y}}})$.

\section{Dataset}
\label{sec:dataset}

\subsection{Overview}
\label{ssec:dataset_overview}
In the realm of instruction-guided voice conversion, indispensable data elements include source speech samples, instructions, and target speech samples. Source speech denotes the original, unmodified vocal expression, whereas target speech refers to the altered vocal output post-stylistic adaptation guided by instructions. Recognizing a deficit of such datasets that incorporate text instruction, we seek inspiration from the InstructSpeech dataset proposed by PromptTTS \cite{guo2023prompttts} and assemble a bespoke dataset to meet our precise requirements.
The dataset is publicly available for further research\footref{fn:asru2023-demo} and samples of the data can be found on our demo website\footnote[2]{\label{fn:asru2023-dataset}{\scriptsize\url{https://text-guided-vc.github.io/asru2023_dataset}}}.
Subsequently, we provide a detailed explanation of the steps and specifics involved in the creation of our dataset.
% To assemble the pertinent data for our work, we elucidate the subsequent methodologies.
% In the realm of instruction-guided voice conversion, indispensable data elements include source speech samples, instructions, and target speech samples that exemplify the desired style change guided by the instructions.
% Our proposed dataset comprises a rich collection of instruction prompts accompanied by style guides, source speech samples, and their corresponding style-altered speech samples. The source speech, in this context, represents the unaltered original speech, while the target speech corresponds to the transformed speech that has undergone style modification. In order to prepare the necessary data for our work, we present the following approaches.
% The instruction plays a key role in adjusting the tone and emotional delivery of the original speech. 

\subsection{Signal Processing Effect Dataset}
\label{ssec:signal_processing_effect_dataset}
SoX \cite{sox} provides a range of functionalities for applying various effects to sound files. Therefore, we propose using SoX as an intuitive approach to generate text instruction-guided voice conversion datasets. To achieve this, we follow these steps:

First, we construct the dataset based on LibriTTS [18] and employ SoX to transform the source speech from LibriTTS into target speech using 14 diverse effects, including bass, treble, chorus, delay, echo, fade, loudness, repeat, reverb, reverse, tempo, volume, pitch, and contrast. Each effect is associated with a set of manually selected distinct parameters that represent different levels of impact on the source speech.

Second, we generate manual instructions that correspond to various SoX commands. We also consider multiple parameters and utilize different adverbs to showcase the diverse levels of effects effectively. For example, consider the ``vol'' (volume) command in SoX. For different values of the ``gain'' parameter, we define corresponding instructions that describe the desired impact on the volume of the source speech. When the ``gain'' parameter is set to 0.5, we define the instruction as ``noticeably decrease the volume of the audio''. This indicates a substantial reduction in volume, resulting in a perceptible decrease in audio loudness. On the other hand, when the ``gain'' parameter is set to 0.75, we might define the instruction as ``gently decrease the volume of the audio.'' This suggests a more subtle reduction in volume, resulting in a mild decrease in audio loudness.
Then, to enhance the diversity of our dataset, we use ChatGPT to paraphrase instructions, generating slightly varied yet instructionally equivalent training examples. By leveraging the paraphrasing capabilities of ChatGPT, we can create variations in the wording and phrasing of the instructions while preserving their underlying meaning and instructional content. Examples of the data can be referred to on our demo website\footref{fn:asru2023-dataset}.

Finally, we have successfully created a total of 355,000 paired training samples, which is equivalent to approximately 600 hours of data. Additionally, we have collected a set of 10,000 samples for testing.

\subsection{InstructSpeech Dataset}
% PromptTTS
\label{ssec:instructspeech_dataset}

We leverage the well-established PromptSpeech \cite{guo2023prompttts} dataset, which provides a rich collection of speech samples paired with style prompts. The dataset encompasses five distinct style factors, namely emotion, gender, pitch, speaking speed, and volume. 
The emotion factor spans diverse types, including cheerful, neutral, whisper, sad, and shouting, facilitating an exploration of extensive emotional expressions in synthesized speech.
The gender factor encompasses two categories. The remaining prosody-related factors are classified into three categories: low, normal, and high, differentiated based on degree.
% The emotion factor encompasses distinct types from cheerful, neutral, whisper, sad, and shouting, allowing us to explore a wide range of emotional expressions in synthesized speech. 
% The gender factor consists of two categories. 
% For the remaining prosody-related factors, there are three categories: low, normal, and high, which are divided according to their degree.

In each speech sample, according to its style prompt, the PromptSpeech dataset provides the metadata of the category of distinct style factors along with their corresponding degrees of level, for instance, pitch is low, volume is low, speed is fast, and emotion is sad. Following such metadata and utilizing a commercial text-to-speech (TTS) system\footnote[3]{\label{fn:asru2023-dataset-tts-api}{\scriptsize\url{https://azure.microsoft.com/en-us/products/cognitive-services/text-to-speech/\#overview}}}, we synthesize the speech data. By doing so, we can treat the style prompt as an instruction, and the synthesized speech generated based on the style prompt can be considered as the target speech. As for the source speech sample, we simply set all the style factors to their normal values.

% The emotion factor encompasses distinct types from cheerful, neutral, whisper, sad, and shouting, allowing us to explore a wide range of emotional expressions in synthesized speech. The gender factor consists of two categories. Additionally, we incorporate three categories for the remaining prosody-related factors, namely low, normal, and high, which provide variations in aspects such as pitch, volume, and speaking speed. These categories help us capture different levels of prosodic features and explore their impact on synthesized speech.

% To enhance the diversity of our dataset and increase its capacity, we employ ChatGPT to paraphrase instructions. Meaningful data augmentation allows us to generate additional training examples that maintain the same instructional intent but exhibit slight variations in wording. By training on a diverse dataset that includes various instruction prompts and their associated style transformations, the model learns to comprehend a wide range of instructions and map them to their respective style conversion.

To bolster the diversity of our dataset, we employ ChatGPT to paraphrase instructions, providing slightly varied yet instructionally identical training examples. This allows our model to learn from a wide range of instructions and their associated style conversions, comprehending and mapping a variety of instructions to their respective style transformations.
Examples of the data are available on our demo website\footref{fn:asru2023-dataset}.
Finally, we have a total of 350,000 paired training samples, which corresponds to approximately 600 hours of data. For the testing data, we have 2,500 samples.

% With this method, the model has the capability to process and comprehend a wide range of instructions, understanding their intended meaning and the corresponding style transformation required. 

\section{Experiment}
\label{sec:experiment}

% \subsection{Model Configuration}
% \label{ssec:subhead}

% Both the AR model and the NAR model have the following three settings.

% \begin{enumerate}

%     \item \textbf{Fine-tune the textual language model}. The textual language model has the encoder-decoder architecture. We fine-tune our neural codec language model on the textual language model as our initialization for the voice conversion task. We refer to this setting as ``Text''.
    
%     \item \textbf{Fine-tune the text-to-speech pretrained language model}. This setting is the same as the textual language model while we utilize text-to-speech as the pre-trained task beforehand and followed by fine-tuning our neural codec language model from it. We refer to this setting as ``TTS''.

%     \item \textbf{Train from scratch}. We utilize the same architecture of the above textual language model but without pre-trained weights. We refer to this setting as ``Scratch''.
    
% \end{enumerate}

% We trained the aforementioned three different configurations on both the AR and the NAR models, resulting in a total of nine possible combinations of models. The naming convention for our voice conversion neural codec models is as following example: For TTS-Text, the former for the AR model utilizing text-to-Speech pretrained language model, and later for the NAR model utilizing textual language model.

\subsection{Evaluation Metric}
\label{ssec:evaluation_metric}

% In order to evaluate the effectiveness of our text instruction-guided voice conversion system, we adopt both the automatic evaluation metric and human evaluation. First, for prosody-related factors, such as pitch, volume, and speaking speed, we employ signal processing methods\footnote[2]{\href{https://github.com/JeremyCCHsu/Python-Wrapper-for-World-Vocoder}{https://github.com/JeremyCCHsu/Python-Wrapper-for-World-Vocoder}} to analyze the differences between the source and target speech, and calculate the accuracy of successful conversions according to corresponding instructions. In more detail, we utilize numerical calculations of fundamental frequency to compare pitch, energy calculations to compare volume, and duration calculations to compare speech rate. 

In order to evaluate the effectiveness of our text instruction-guided voice conversion system, we adopt both the automatic evaluation metric and human evaluation. 
First, for prosody-related factors, such as pitch, volume, and speaking speed, we utilize signal processing methods\footnote[4]{\label{fn:asru2023-signal-processing-tool}{\scriptsize\burl{https://github.com/JeremyCCHsu/Python-Wrapper-for-World-Vocoder}}} to compute the differences between the source speech and target speech. 
In the case of pitch, we calculate the average fundamental frequency of each sound frame. For volume, we calculate the average energy of each sound frame in the audio signal. For speaking speed, we directly compare the lengths of the two audio clips.
% For pitch, volume, and speaking speed evaluations, we respectively calculate the fundamental frequency, the average squared sum of each sound frame, and directly compare audio clip lengths.
To measure the accuracy, we compute the proportion of successful conversions out of the total number of testing cases. A successful conversion is defined as one where the target speech has been accurately transformed in accordance with the instructions. For instance, if the instruction is ``speak in a higher pitch'', we calculate the fundamental frequency of both the source and converted speech. We compare whether the converted speech has a higher frequency, which indicates a higher pitch, aligning with the instructions.
In addition, we utilize the NISQA \cite{mittag21_interspeech} (Speech Quality and Naturalness Assessment), which is a deep learning framework for speech quality prediction, to evaluate the quality of the synthetic speech automatically.

% However, when it comes to evaluating the emotion factor, signal processing methods alone may not provide accurate results. Furthermore, there is currently no effective automatic method to evaluate the consistency between the given instruction and the resulting target speech in terms of style.
% Therefore, we perform mean opinion score (MOS) evaluation and ask participants to investigate whether the descriptions of emotion provided in the instructions have both significant and correct effects on the resulting target speech, thereby determining the extent to which the influence is evident. In detail, we randomly select a total of 30 utterances to evaluate the performance of models under different configurations. Each sentence will be rated by at least five individuals.

% However, evaluating the emotional aspect presents challenges as signal processing evaluation methods alone may not yield precise outcomes. Additionally, an effective automatic methodology to ascertain the congruence between the provided instruction and the resultant target speech in terms of style is currently absent. 

However, evaluating the emotional aspect poses challenges as signal processing evaluation methods may not provide precise results. Furthermore, a robust automatic method for assessing the congruence between given instruction and resultant speech style is currently lacking.
Consequently, we turn to employing the MOS evaluation, tasking participants with verifying the effect and correctness of especially non-prosodic instructions, for instance, emotion, echo, and reverberation effects, on the resulting target speech to gauge the level of evident influence. In general, employing human evaluation not only enables the assessment of speech quality, but in our context, it also serves as a tool to measure and evaluate the consistency between the textual narratives in the instructions and the stylistic presentation in the target speech. In the detail of MOS evaluation, we randomly select a total of 30 utterances to assess the performance under various model configurations. Each utterance is evaluated by at least five participants.

\begin{table}[t]
    \centering
    \resizebox{.35\textwidth}{!}{
    \begin{tabular}{l|ccc|c}
    \toprule
        Setting & \begin{tabular}[c]{@{}c@{}} Pitch\end{tabular} & \begin{tabular}[c]{@{}c@{}}Volume\end{tabular} & Speed & Mean \\ 
        \midrule
        % \cline{1-5}
        % \rule[0.05ex]{\linewidth}{1pt}
        TTS-TTS & 70.82 & 86.73 & 84.61 & 86.73 \\
        TTS-Text & 71.39 & 86.78 & 84.61 & 86.78 \\
        TTS-Scratch & 71.44 & 76.78 & 86.91 & 76.78 \\
        Text-TTS & 69.61 & 87.43 & 89.72 & 87.43 \\
        Text-Text & 69.72 & 88.50 & 89.72 & 88.50 \\
        Text-Scratch & 70.19 & \textbf{88.60} & \textbf{89.78} & \textbf{88.60} \\
        Scratch-TTS & \textbf{72.75} & 76.51 & 86.91 & 76.51 \\
        Scratch-Text & \textbf{72.75} & 78.92 & 86.91 & 78.92 \\
        Scratch-Scratch & 72.23 & 78.54 & 86.91 & 78.54 \\
        \bottomrule
        \end{tabular}
    }
    \caption{Accuracy (\%) of prosody-related results.}
    \label{automatic-metric}
\end{table}

\begin{table}[t]
    \centering
    \resizebox{.45\textwidth}{!}{
    \begin{tabular}{l|cc|c}
    \toprule
        Setting & \begin{tabular}[c]{@{}c@{}} Signal Procssing Effect\end{tabular} & \begin{tabular}[c]{@{}c@{}}InstructSpeech\end{tabular} & Mean \\
        \midrule
        TTS-TTS & \textbf{2.36} & \textbf{2.75} & \textbf{2.56} \\
        TTS-Text & 2.29 & 2.56 & 2.43 \\
        TTS-Scratch & 2.22 & 2.63 & 2.43 \\
        Text-TTS & 2.24 & 2.74 & 2.49 \\
        Text-Text & 2.24 & 2.54 & 2.39 \\
        Text-Scratch & 2.22 & 2.60 & 2.41 \\
        Scratch-TTS & 2.24 & 2.67 & 2.46 \\
        Scratch-Text & 2.24 & 2.47 & 2.36 \\
        Scratch-Scratch & 2.24 & 2.52 & 2.38 \\
        \midrule
        Ground Truth & 3.04 & 3.86 & 3.46 \\
        \bottomrule
        \end{tabular}
    }
    \caption{Estimated results of MOS using NISQA. The setting of ground truth refers to testing on the ground truth of target speech.}
    \label{MOS-NISQA}
\end{table}

\begin{table}[t]
    \centering
    \resizebox{.35\textwidth}{!}{
    \begin{tabular}{l|cc}
    \toprule
        Setting & \begin{tabular}[c]{@{}c@{}} Quality\end{tabular} & \begin{tabular}[c]{@{}c@{}}Instruction\end{tabular} \\
        \midrule
        TTS-TTS & {2.78 \(\pm\) 0.20} & {3.48 \(\pm\) 0.19} \\
        Text-Text & {2.75 \(\pm\) 0.16} & {3.39 \(\pm\) 0.17} \\
        Text-Scratch & \textbf{2.81 \(\pm\) 0.14} & \textbf{4.19 \(\pm\) 0.14} \\
        \midrule
        Ground Truth & {3.76 \(\pm\) 0.17} & {3.64 \(\pm\) 0.19} \\
        \bottomrule
        \end{tabular}
        }

    \caption{The chart represents the MOS evaluation results, presented with 95\% confidence intervals. The ``Quality'' field requests participants to evaluate the quality of speech, disregarding vocal effects, while the ``Instruction'' field solicits participants to assess the stylistic consistency between the instructions and the target speech.}
    % \caption{The chart depicts MOS results, where "Quality" pertains to speech quality assessment excluding vocal effects, and "Instruction" evaluates the stylistic consistency between instructions and target speech.}
    
    \label{human_evaluation}
\end{table}

% Specifically, we conduct the mean opinion score (MOS) to evaluate the speech quality from the human perspective. Our evaluation criteria focus on assessing the alignment between the expression of speech and the given instruction. We aim to determine how well the generated speech conforms to the content and style specified in the instruction.

% Therefore, we train classifiers specifically designed to categorize and identify different emotions present in the output speech. These classifiers enable us to assess the accuracy of the emotion representation and ensure that the converted speech conveys the intended emotional content. 

% By combining signal processing techniques for prosody-related factors and trained classifiers for emotion factors, we can automatically evaluate the accuracy and fidelity of the style transformations achieved by our text instruction-guided voice conversion system. This evaluation process allows us to measure the effectiveness of the system in accurately capturing and producing the desired style attributes specified in the instructions. 

% Finally, we conduct the mean opinion score (MOS) to evaluate the speech quality from the human perspective. Our evaluation criteria focus on assessing the alignment between the expression of speech and the given instruction. We aim to determine how well the generated speech conforms to the content and style specified in the instruction. \\

% {\bf Human evaluation:} In order to

\subsection{Result}
\label{ssec:result}

The naming convention for our voice conversion neural codec models is as following example: For TTS-Text, the former for the AR model utilizing text-to-speech pre-trained language model, and later for the NAR model utilizing textual language model.

First, we explore the performance of different models in prosody-related aspects. In table \ref{automatic-metric}, the configuration of Text-Scratch exhibits the best average performance. Fine-tuning the model from the textual language model allows for a better understanding of the content of instructions, enabling the model to perform speech transformations well according to the given instructions.

% compare Text-Text to Text-Scratch.

Second, we conduct the experiment on speech quality prediction using NISQA. We test on our generated target speech for Sox and InstructSpeech datasets respectively. In table \ref{MOS-NISQA}, the configuration of TTS-TTS yields the best prediction of mean opinion score (MOS) for both datasets. The result demonstrates that the model benefits from finetuning from the text-to-speech pre-trained language model, which can synthesize the speech with better quality.

Finally, we examine whether fine-tuning using a textual model assists in improving the understanding of instructions and whether fine-tuning with a TTS model augments the quality of synthesized speech by human evaluation. 
Consequently, we select several settings for analysis, which include Text-Text, TTS-TTS, Text-Scratch, and Ground Truth. Among these, Text-Scratch exhibits superior performance in prosody-related results, potentially signifying enhanced capabilities to comprehend instructions. TTS-TTS renders the best performance in estimated MOS results. Text-Text, in contrast, is intended for comparison with Text-Scratch, with both AR and NAR utilizing the textual model for fine-tuning simultaneously. Finally, Ground Truth serves as a topline for comparison. In experimental detail, we employ MOS with two distinct criteria for evaluation. The first criterion requests participants to use the source speech as a benchmark, assessing whether the effects manifested in the target speech align with the textual description provided in the instructions. This criterion requires scoring based on the effects observed in the target speech, disregarding speech quality. The second criterion asks participants to rate the target speech based solely on speech quality, ignoring vocal effects. The result is shown in Table \ref{human_evaluation}.

In Table \ref{human_evaluation}, analyzing the results from the aspect of stylistic consistency between the instructions and the target speech, the Text-Scratch configuration showcases an impressive understanding of instructions, successfully translating corresponding stylistic directives into the target speech. This substantiates the feasibility of text-instruction-guided voice conversion. Comparatively, the Text-Text configuration, although fine-tuned using the textual language model for both the AR model and the NAR model, renders a relatively weaker performance. From this, two conclusions can be inferred. First, the AR model can benefit from fine-tuning using the textual language model, enabling the model to possess a proficient understanding of text semantics. Second, for the NAR model, fine-tuning with the pre-trained textual language model seems to act as an impediment rather than an aid. On the other hand, the Text-Scratch configuration outperforms the ground truth by a significant margin. We attribute this to certain effects that are less prominent in the original datasets and are used as ground truth, but became more conspicuous in the generated speech due to the model having learned from extensive data, thus garnering higher evaluation scores.
From the perspective of speech quality, results reveal that the configuration that obtains the highest estimated MOS, TTS-TTS, is not the top scorer. This suggests a discrepancy between human evaluation outcomes and model-based predictions. 
In conclusion, the evaluation results showcase our dazzling text comprehension ability of the model, effectively translating instructions into corresponding styles presented in the target speech. Nonetheless, the quality of the speech, when compared, did not equally impress, indicating our future efforts will need to focus on improving speech quality.

\subsection{Analysis}
\label{ssec:analysis}
% \textbf{Analysis of generalizability}
% Finally, in order to access the generalization and robustness capabilities of the model, we carry out experiments to evaluate our model on out-of-domain instructions that are never seen during training. We accomplish unseen instructions by amalgamating distinct types of instructions, one exclusively appearing in the Signal Processing Effect Dataset and the other solely in the InstructSpeech Dataset.

First, to gauge the capacity of the model for generalization and robustness, we implement experiments designed to assess its performance with previously unseen, out-of-domain instructions. These novel instructions are assembled by incorporating unique directive types, some derived exclusively from the Signal Processing Effect Dataset, while others originate solely from the InstructSpeech Dataset. 
% For example, we combine the instructions related to reverberation (``add reverberation to the audio and say in whispering style.''), which are only present in the former dataset, with the instructions related to emotion since emotion-related information only appears in the later (``say in sad tone and enhance the speech's spatial dimension and depth''). 
For instance, we meld instructions associated with reverberation from the former dataset, such as ``add reverberation to the audio'', with emotion-related instructions from the latter dataset like ``say in a whispering style''. Such combination produces ``add reverberation to the audio and say in whispering style''. We dissect the different elements of the instruction and observed their individual effects as well as their combined effect on the source speech. Examples are available on our demo website\footref{fn:asru2023-demo}.

Second, for analysis of adverbial modifiers, we look into the effect of different degrees of adverbs in the instructions on the generated results. In addition, we include instructions without adverbs for further observation. The following are some examples, 
    1) Decrease the speed of the audio \textbf{slightly}. 
    2) Decrease the speed of the audio. 
    3) Decrease the speed of the audio \textbf{notably}. 
    4) Decrease the speed of the audio \textbf{extremely}. 
    % In addition, we also investigated the effect of different adverbs when combined with emotion words on the resulting speech. For example, 
    Moreover, we examined how the combination of different adverbs and emotional words influences the resulting speech. For example,
    1) A girl speaking in a \textbf{slightly} sad tone. 
    2) A girl speaking in a sad tone. 
    3) A girl speaking in an \textbf{extremely} sad tone.
% Demonstrations are available here\footnotemark[1].
Audio samples are available on our demo website\footref{fn:asru2023-demo}.

\begin{figure}[H]
    \centering
    \includegraphics[width=0.55\textwidth]{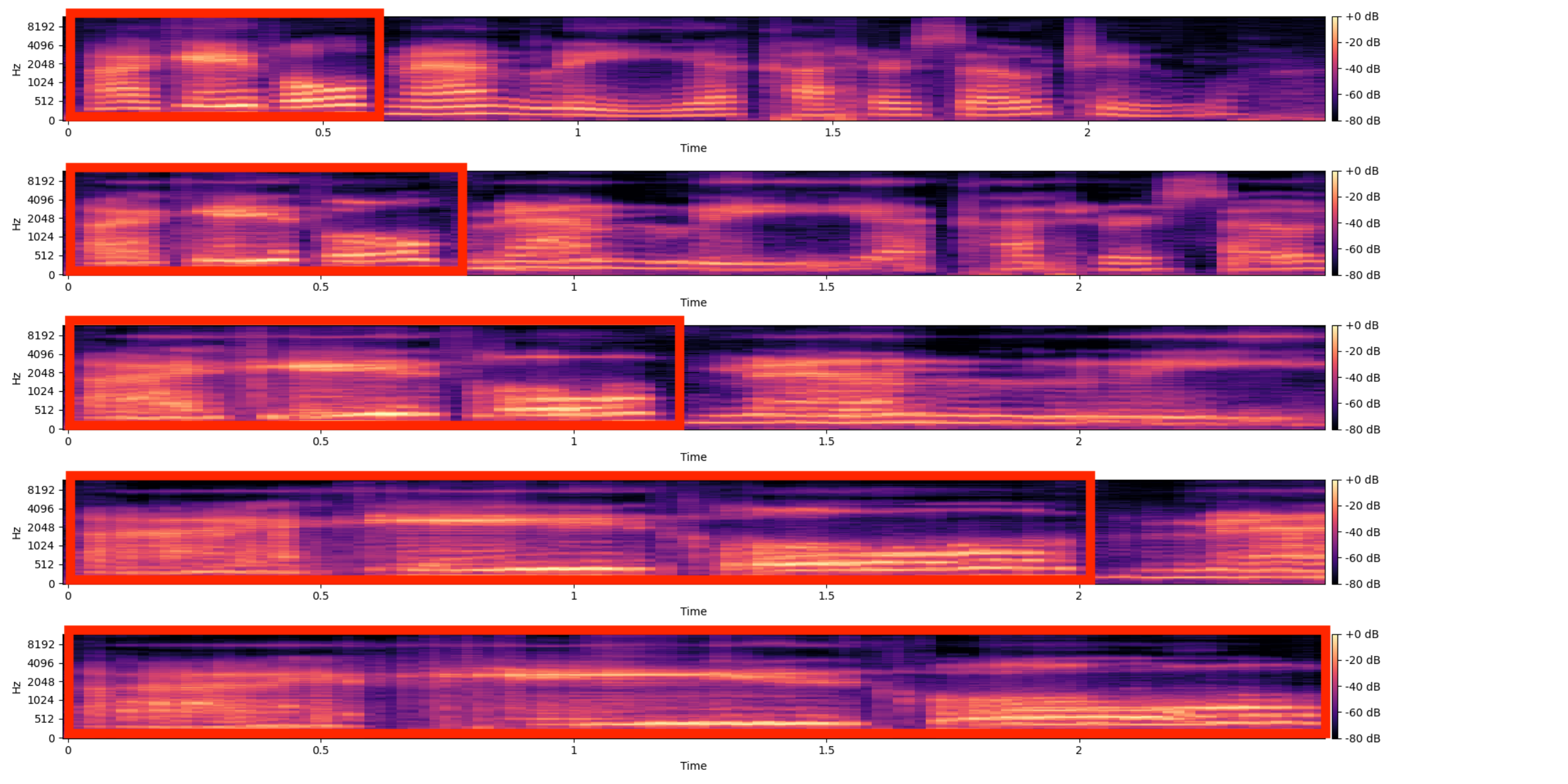}
    
    \caption{From top to bottom, (1) Spectrogram of the source speech (2) Spectrogram of the target speech with the instruction as ``Decrease the speed of speech slightly.'' (3) ``Decrease the speed of speech.'' (4) ``Decrease the speed of speech notably.'' (5) ``Decrease the speed of speech extremely.'' The length of the audio, originally 2.47 seconds in the source speech, is extended to 3.17, 5.25, 8.44, and 10.27 seconds respectively.}
    
    \label{adverb}
\end{figure}

Our evaluation criteria focus on assessing whether different degrees of adverbs have varying degrees of effect on the results. 
In Fig. \ref{adverb}, it is evident that adverbs of different degrees indeed have a noticeable and varying effect on the source speech. We align the time axes for easy comparison and use red boxes to denote prominent spectral features. As can be seen, the speech signal exhibits varying degrees of slowdown effects in accordance with the magnitude of the adverb used.
Therefore, the model has learned to associate the differences in adverbs with varying degrees of style transformation.
% Examples in Fig \ref{reverberation-whispering} and Fig \ref{adverb} are available on our demo website\footnote[5]{ \href{https://asru2023taipei.github.io/demo}{https://asru2023taipei.github.io/demo}\label{fn:demo}}. 
Examples in Fig \ref{adverb} are available on our demo website\footref{fn:asru2023-demo}.
% \footnote[5]{ \href{https://asru2023taipei.github.io/demo}{https://asru2023taipei.github.io/demo}\label{fn:demo}}. 
In addition, due to space constraints, not all results are presented within the article. Please refer to our demo website\footref{fn:asru2023-demo} for a detailed view of all results.

% Examples in Fig \ref{reverberation-whispering} and Fig \ref{adverb} are available here\footnote[5]{See \href{https://asru2023taipei.github.io/demo}{\color{blue} here} for demonstrations of our work. \label{fn:demo}}. In addition, due to space constraints, not all results are presented within the article. Please refer to our demo webpage for a detailed view of all results.
% footnote[3]{See \href{https://asru2023taipei.github.io/demo}{\color{red} here} for demos of our work.}

% \href{https://asru2023taipei.github.io/demo/}{here}.

% \href{https://github.com/JeremyCCHsu/Python-Wrapper-for-World-Vocoder}{https://github.com/JeremyCCHsu/Python-Wrapper-for-World-Vocoder}

% \begin{figure}[H]
%     \centering
%     % \includegraphics[width=0.5\textwidth]{figures/generalization.png}
%     \includegraphics[width=0.55\textwidth]{figures/out-domain-demo.pdf}

%     \caption{From top to bottom, (1) Spectrogram of the source speech (2) Spectrogram of the target speech with the instruction as "Add reverberation to the audio." (3) "Say in whispering style." (4) "Add reverberation to the audio and say in whispering style."}

%     \label{reverberation-whispering}
% \end{figure}

\section{Conclusions and future works}
\label{sec:conclusion_future_works}

% In this paper, we explore the potential of text-intruction-guided speech-to-speech voice conversion. We leverage natural language instructions without any constraints on the specific format to control versatile aspects of speech generation within a unified framework. We have observed that our model demonstrates a strong ability to comprehend instructions and successfully applies style transformations to the source speech, which confirms the feasibility of text-instruction-guided voice conversion.
% In the future, both enhancing the quality of the synthesized speech and increasing the diversity and richness of the instruction content could be directions for further improvement.

In this paper, we have explored the potential development of text-instruction-guided voice conversion. Within a unified framework, we leverage natural language to guide the stylistic transformation in voice conversion. 
This not only fully harnesses the richness of language to increase the flexibility of voice conversion tasks, but it also avoids the limitations of any standardized instruction formats or the necessity of reference speech in traditional voice conversion.
From our experimental results, we observe that our model has a remarkable ability to understand text instructions and successfully transforms styles according to these instructions. This demonstrates the feasibility of text-instruction-guided voice conversion and its excellent potential for further development.
In the future, we will focus on improving the quality of speech and increasing the diversity and richness of instruction types.
% In the future, we will aim to concentrate on improving speech quality and expanding the diversity and richness of instruction types present in our dataset.

\section{Acknowledgements}
This research is based on work that is funded by a 
Googler-Initiated Grant from Shuo-yiin Chang. 
Additionally, we thank the National Center for High-performance Computing (NCHC) of National Applied Research 
Laboratories (NARLabs) in Taiwan for providing 
computational and storage resources.

\bibliographystyle{IEEEbib}
\bibliography{strings,refs}

\end{document}